\documentclass[11pt]{article}
\usepackage{amssymb}
\usepackage{amsmath}
\usepackage[dvips]{epsfig}

\def\##1{\underline #1}
\def\=#1{\underline{\underline #1}}
\def\.{\mbox{ \tiny{$^\bullet$} }}

 \def\wS{w_s}
 \def\wF{w_f}
 \def\gS{\gamma_s}
 \def\hS{h_{sw}}
 \def\chiV{\chi_v}
 \def\deg{^\circ}
 \def\dSW{\delta_{sw}}

\def\c#1{\cite{#1}}

\bigskip

\begin{document}

\noindent{\Large {\bf Blending of nanoscale and microscale in uniform large--area sculptured thin--film architectures} }
\vskip 0.4cm

\noindent{\bf Mark W. Horn, 
 Matthew D. Pickett, 
Russell Messier
and
Akhlesh Lakhtakia\footnote{Corresponding author. Tel: +1 814 863 4319; Fax: +1 814 865 9974;
E--mail: AXL4@psu.edu}}

\vskip 0.2cm
\noindent{\em 
Department of Engineering Science \& Mechanics,
Pennsylvania State University, \\
University Park, PA 16802--6812, USA}

\vskip 0.5cm

\noindent {\bf Abstract} \\
The combination of large thickness ($>3$~$\mu$m), large--area uniformity
(75~mm diameter), high growth rate (up to 0.4~$\mu$m/min) in 
 assemblies of complex--shaped nanowires on lithographically
defined patterns has been achieved for the first time. The nanoscale and the microscale
have thus been blended together in sculptured thin films with transverse architectures.
SiO$_x$ ($x\approx 2$) nanowires were grown by electron--beam evaporation onto
silicon substrates both with and without photoresist lines (1--D arrays) and checkerboard
(2--D arrays) patterns. Atomic self--shadowing due to oblique--angle deposition enables
the nanowires to grow continuously, to change direction abruptly, and to maintain
constant cross--sectional diameter. The selective growth of nanowire assemblies on
the top surfaces of both 1--D and 2--D arrays can be understood and predicted
using simple geometrical shadowing equations.




\section{Introduction}
Raise the temperature and lower the pressure sufficiently, and virtually any solid 
material shall begin to evaporate. Condense the directional vapor on a substrate at a temperature 
less than about a third of its melting point, and you will obtain a columnar thin film 
with column diameters on the order of tens of nanometers.  The technology of  
columnar thin films (CTFs) was born in 1885, when Kundt collected the vapor from a solid on a 
substrate held at an oblique angle to the  incident, directed vapor flux \c{1}.  Even before the commercial 
advent of transmission electron microscopes (TEM) in the 1950Õs and scanning electron microscopes (SEM) 
in the 1960Õs, it was known from optical experiments that such films are anisotropic \c{2}~---~not unlike crystals.  
However, unlike the anisotropy of crystal structures, the anisotropy of CTFs is 
morphological: Self--shadowing at the atomic level \c{3} leads to parallel
columns growing towards the obliquely
incident vapor flux, albeit at an angle to the average direction  \c{4}. TEMs and SEMs eventually 
provided visual evidence of the nm--scale morphology of CTFs grown by oblique--angle deposition \c{5}, 
thereby confirming the optical analogy with crystals.

Also unlike crystals, the separated and slanted nanocolumns 
(currently termed {\em nano\-wires\/}) can be shaped during growth by substrate motion \c{6}.  
Elementary experiments were 
reported by (i) Young and Kowal in 1959 \c{7} who rotated the substrate about an axis 
passing normally through it, and (ii) Nieuwenhuizen and Haanstra in 1966 \c{5}
 who altered the substrate tilt with respect to the average direction
 of the incident vapor flux once during deposition.  The products of these 
 experiments were the precursors of sculptured thin films (STFs) developed 
 during the last decade \c{6}, chiefly for optical  but also other applications \c{8}.  
 
STFs are assemblies of parallel, shaped nanowires that can be fabricated by 
 design using physical vapor deposition techniques, such as thermal and arc evaporation, 
 sputtering, and pulsed laser ablation \c{4,8,9,10}.  As the nanostructure comprises 
 multimolecular clusters of 3--5 nm diameter, rapid changes in the average direction of the 
 incident vapor flux relative to the substrate lead to the growth of parallel nanowires of 
 curvilinear shapes. At infrared, visible and lower ultraviolet wavelengths, the 
 assembly of nanowires can be effectively considered as a continuous 
 anisotropic medium whose electromagnetic response properties are 
 inhomogeneous normal to the substrate \c{11}~---~which allows the 
 exploitation of commonplace design techniques for devices such as optical polarizers, filters and sensors \c{8,9}.
 
The STFs fabricated thus far possess, for the most part, simple transverse architectures, 
 as the constituent nanowires are randomly nucleated on the substrate. The film
 growth rates can also be low, depending on the deposition method employed,
 due to the large oblique angles during deposition.
 Furthermore, the transverse area rarely exceeds 1~cm~$\times$~1~cm without significant loss of 
 transverse uniformity.  These three factors have stymied the economic exploitation of 
 STFs, despite the theoretical design and experimental realization of many STF--based 
 optical devices \c{8,9}. Fabrication of complex STFs of large transverse area 
 and at high deposition rates would increase the economic attraction of STFs.
 
We present here a newly realized technique to economically grow thick, uniform, 
 large--area STFs with transverse architectures that blend the nanoscale and the 
 microscale.  Our work goes beyond initial work on patterned growth in thin
 films \c{12}--\c{14} in 
 that we examine different microscale architectures
 employing a range of feature sizes and shapes with controlled depth and shape of 
 topography, and have achieved this over relatively large ($75$~mm dia) areas.
 
We have chosen to illustrate the growth of sculptured nanowire assemblies by 
 examining the deposition of SiO$_x$, a dielectric substance, on substrates
 photo\-litho\-graphically  patterned 
 with posts, 
 holes, and various densities of lines and spaces that were generated
using a deep 
 ultraviolet (DUV) stepper.  Although we have examined the deposition of several materials
 (such as SnO$_2$, Mo, Cr and Al) in addition to SiO$_x$,
  for the sake of brevity we only 
 show examples of SiO$_x$ films deposited at a single angle of vapor--flux incidence.  
We continue to investigate STFs grown
 at various oblique angles of vapor--flux incidence, deposition rates, and substrate 
 rotational velocities in order to better understand how the initial topography of the 
 substrate affects the growth and expansion of nanowires.
 
\section{Experimental}

A 10--kW dual electron--gun evaporator was used to obliquely deposit material  on 
either fixed or rotating substrates, as shown schematically in Fig. 1a and b.  The system was 
cryopumped to a base pressure of $5\times10^{-7}$ Torr.  A 100--cc, 50--mm diameter
crucible was charged with pellets of SiO$_2$ before each deposition.  
Deposition rates were measured using a quartz crystal monitor (QCM) held perpendicular
to the vapor flux. The measured deposition rate depends on the mass density of
the vapor flux, and is quite different from the film growth rate defined as the
rate of increase of  film thickness.
Deposition rates from 54 to 480~nm/min 
 were measured, while the film growth rates varied
 between 100 and 1000~nm/min. Nanowire shapes were controlled
 {\em via\/} computer, and
substrate rotational velocities ranged from 0.5 to 36~rpm.

\begin{figure}[!ht]
\centering \psfull \epsfig{file=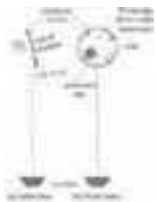}
\caption{Schematic of the deposition system. (a) Side view; (b) Front view.
}
\end{figure}

The throw distance to the center of the rotating  substrate was 250~mm, 
and the center of the wafer was nominally positioned directly above the center of the crucible; see Fig. 1a.  
Unlike previous work, the electron beam in our system was rastered over a large portion of the crucible, 
and the flux produced more closely resembles a uniform flat source, analogous to a 
plane wave in electromagnetics textbooks \c{15}, than a point source. Furthermore, the 
mean free path during deposition was much greater than the source--to--substrate distance, 
leading to line--of--sight deposition. Thus, the arriving vapor flux is highly directed, 
thereby eliminating the need for collimation.  For example, we observed uniformity of 
thickness and morphology of complex sculptured nanowires across the diameter of a 
75~mm substrate.  However, highly directed vapor flux leads to 
significant  local nonuniformity due to shadowing effects for oblique incidence on non--planar
(i.e., patterned) substrates. 

The substrates upon which the sculptured nanowire assemblies were deposited consisted of a 
1~cm~$\times$~1~cm die of a silicon wafer 
patterned with a photoresist stack placed on top of and near the outer edge of a 75~mm silicon wafer, 
as shown schematically in Fig. 1b.  The substrates were clipped into place on a 
rotating substrate holder driven by an external stepper motor via 
a rotary vacuum feed--through.  For all films presented in this paper, 
the vapor flux was directed at an angle
$\chiV=15^\circ$ to the substrate plane.  

The photoresist features shown 
throughout this paper were generated using a resolution test mask and a DUV (248~nm) stepper.  
A Shipley UV5 photoresist was applied to a 150~mm silicon wafer 
coated with a Shipley lift--off photoresist (LOR) layer.  The two--layer photoresist stack~---~used
to facilitate selective removal of a deposited film~---~was
then exposed, baked after exposure,  and then developed to yield a wide variety of 1--D and 2--D 
features from 0.4~$\mu$m to hundreds of microns  wide and 0.8~$\mu$m high with nearly vertical sidewalls.  
After deposition, the films were cleaved into smaller pieces and were analyzed in cross--section using a LEO 1530 
field emission SEM (FESEM), usually at voltages between 2 and 5~kV to minimize charging of the dielectric nanowires.  

\section{Results}
Figures 2a and b are cross--sectional SEM micrographs of a two--section STF comprising helical (chiral) 
nanowires and chevronic (nematic) nanowires grown sequentially, 
without stopping the evaporation process. A flat silicon wafer with no 
photoresist pattern was used. 
The substrate revolution rate for the  chiral section (with 3--D morphology) of the STF was 4~rpm.  
The nematic section (with 2--D morphology) of the STF went through nine rapid changes of the 
substrate tilt ($+75^\circ$ to $-75^\circ$ with respect to the normal to the substrate), 
one change every 30~s.  The deposition rate was fixed at 120~nm/min.  The total height  (i.e., thickness) of the film, as 
measured using the SEM micrographs, is approximately 4~$\mu$m.  
The total deposition time was 10 min, yielding an average
growth rate of 400~nm/min.   Fig.  2a shows a total cross--section of the  film,
whereas Fig. 2b shows a high magnification of the region where the film 
abruptly, yet continuously, transitions from the chiral section to the nematic section. 

\begin{figure}[!ht]
\centering \psfull \epsfig{file=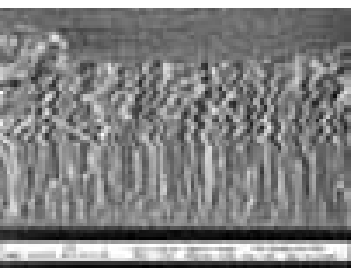}
 \epsfig{file=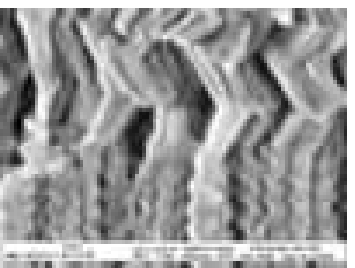} \\
(a) \hskip 1.25 in (b)
\caption{(a) Cross--sectional SEM micrograph of a two--section STF comprising a nematic
section on top of a chiral section, grown without turning off the vapor flux. The film
was deposited on a flat silicon (Si) substrate. (b) Highly magnified view of the region
where the film transitions from the 2~$\mu$m--high chiral section to the 
1~$\mu$m-high nematic section. Note the
continuity between the helical and the chevronic nanowires.
}
\end{figure}

Figures 3a and b are cross--sectional SEM micrographs of a two--section STF 
taken at the center and edge, respectively, of a 75~mm substrate cleaved along its diameter.
This 2--section chiral STF was grown by varying the deposition rate
while holding the substrate rotational velocity fixed at approximately 2~rpm.  
The deposition rate was set at 54~nm/min for the first section, and increased to 
about 300~nm/min for the second section.   The structure and density of the 
nanowires are very similar in both micrographs, even though 
 the center of the substrate must receive a relatively constant vapor flux 
during an entire revolution while  the edge must experience 
some variation.  
In our deposition chamber, if there was a large variation in deposition rate either from the 
bottom  to the top of the substrate or from the left to the
right of the substrate (see Fig. 1a),   a 
significant variation in the chiral morphology of the nanowires should be expected. 
However, at least at the nanoscale, we see very little difference in the nanowire 
assemblies deposited at the edge of a 75~mm rotating substrate with those grown 
at its center, which is consistent with our assumption that the  wide melt pool
in the crucible is 
acting more like a planar deposition source than a point source. 

\begin{figure}[!ht]
\centering \psfull \epsfig{file=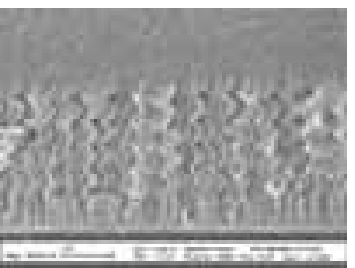}
 \epsfig{file=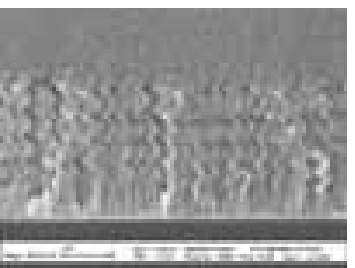} \\
(a) \hskip 1.25in (b)
\caption{Cross--sectional SEM micrographs of a two--section STF comprising two different
chiral sections, grown on a flat 75~mm dia silicon substrate without turning off the vapor flux. 
 (a) Center of the substrate;
(b) Edge of the substrate. The lower section is 0.75~$\mu$m
high, and the upper section is about 2.25~$\mu$m high.
Note the uniformity in morphology in film growth over a large
substrate.
}
\end{figure}

Figures 4a and b are cross--sectional SEM micrographs of a yet another different 
STF taken at the center and an edge, respectively, of a 75~mm dia silicon substrate.  
This 2--section film was grown by fixing the deposition rate at 
120~nm/min and abruptly changing the substrate rotation velocity from 
1 to 36~rpm.  Notice again the similarity in morphology in the two different 
cross--sections, confirming the high degree of uniformity of deposition across large areas.  
Also notice that, as the substrate rotational velocity is increased, the helical shape of the 
individual nanowires is lost and larger diameter nanocylinders  grow orthogonal 
to the substrate surface. 

\begin{figure}[!ht]
\centering \psfull \epsfig{file=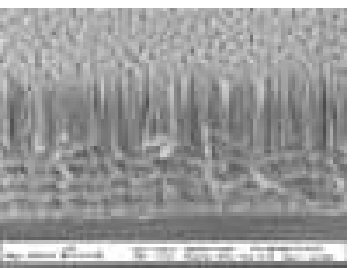}
 \epsfig{file=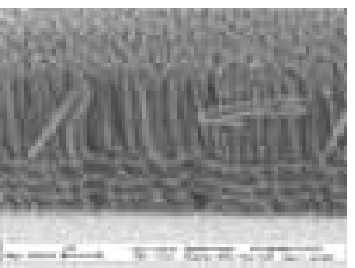} \\
(a) \hskip 1.25in (b)
\caption{Cross--sectional SEM micrographs of a two--section STF comprising two different
chiral sections, grown on a flat 75~mm dia silicon substrate without turning off the vapor flux.   (a) Center of the substrate;
(b) Edge of the substrate. Note the uniformity in morphology in film growth over a large
substrate.
}
\end{figure}

Figure 5 is a cross--sectional SEM micrograph of a STF grown simultaneously
with the two--section STF of Fig. 2, except that the former was deposited on a 
photoresist--patterned substrate. The  micrograph in Fig. 5 shows the growth behavior 
at a   single 1--D  topographical step. The nanowire assembly in 
the left portion of Fig. 5 grew on top of the 0.8--$\mu$m high photoresist 
and is morphologically similar to the STF of Fig. 2. Likewise, the morphology of the nanowire
assembly   on the far right of Fig. 5,  well away from the step, is essentially the same. In all these cases 
there is no shadowing due to the substrate; only self--shadowing occurs at the atomic level 
that leads to the growth of individual nanowires. The film that grows on the side of the photoresist 
step appears much denser and devoid of distinct nanowires. However, zone 1 
(see Fig. 5) shows indications of layered, shadowed--growth during the 
3--D chiral--section growth, and zone 4 displays oblique columnar shadowed-growth 
related to the 2--D nematic section. Zones 2 and 3, farther away from the shadowing ledge, 
have the appearance of chiral section. The shadowing distance of a 
0.8--$\mu$m high step is about 3~$\mu$m to the right of the step (zone 2), followed by about a 
1~$\mu$m transition region (zone 3) to the un-shadowed region on the right of the micrograph.  

\begin{figure}[!ht]
\centering \psfull \epsfig{file=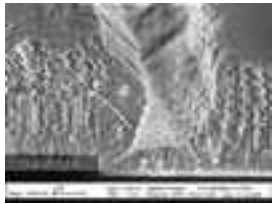}
\caption{Cross--sectional SEM micrograph of a two--section STF comprising a nematic
section on top of a chiral section, grown without turning off the vapor flux. The film
was deposited on a flat silicon   substrate with a 0.8--$\mu$m--high photoresist line. 
}
\end{figure}

The STF of Fig. 3 was also grown on a patterned--die substrate with  
1.5--$\mu$m--wide and 0.8--$\mu$m--high photoresist lines. Figures 6a and b
provide cross--sectional SEM micrographs of this film. In Fig. 6a, we have focused on the morphology
developed over a
single line.
 Shadowing by the 1--D topography  is evident on both sides of the line. To the
 left of the line, the shadowed--morphology zones are similar to
 those in Fig. 5.
 At a distance far enough 
 to the left of the line 
 such that no shadowing  occurs, the morphology and thickness of the STF on the flat Si substrate 
 are the same as that grown on top of the photoresist line. In the center of the micrograph, but
 still to the  left of photoresist line, 
 the film is noticeably thinner due to shadowing by the line, but it
 exhibits a similar chiral morphology. Near the line, the shadow--related morphology can be
 divided into several zones. Zone 1 is related to shadowed--growth on the lineÕs sidewall,
  while zone 2 is shadowed--growth emanating from the Si substrate.  
  The boundary between these two zones appears to follow a power--law growth evolution that 
  results from competition between growths on the sidewall and the bottom surface \c{4}. 
  The boundary between zone 2
  and  the transition zone 3 (at about 4~$\mu$m from the left sidewall
  of the line) is less abrupt than in Fig. 5 (where the upper section is of the
  nematic type in which the shadowing effects are more significant, as  discussed in the next section). 
 The unshadowed--growth at the left is almost 5.5~$\mu$m from the left sidewall of the
  line.
  
  The very thin layer at the far right of the micrograph in Fig. 6a is shadow--related growth onto an 
  adjacent space between
  two photoresist lines with sidewall separation of 1.5~$\mu$m; see Fig. 6b. 
  A thin but visible coating grew on the sidewalls and the bottom of both 1.5--$\mu$m--wide 
  spaces between adjacent
  photoresist lines. A lower magnification, edge--on view of this film is provided in Fig. 7a.

\begin{figure}[!ht]
\centering \psfull \epsfig{file=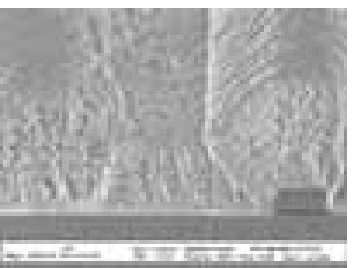}
 \epsfig{file=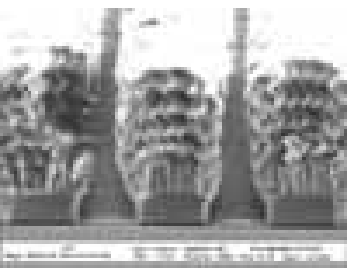} \\
(a) \hskip 1.25in (b)
\caption{Cross--sectional SEM micrographs of a two--section chiral STF deposited, on
a patterned--die Si substrate with 1.5--$\mu$m--wide and 0.8--$\mu$m--high
photoresist lines, during the same experiment as the film presented in Fig. 3. Examples
of shadow--controlled
morphology are shown: (a) at a sidewall, and (b) between adjacent lines.
}
\end{figure}

Figures 7a--c are additional cross--sectional SEM micrographs of the same sample
as shown in Fig. 6, but taken edge--on at 
different locations along a 1--D array of photoresist lines. Fig. 7a focuses on growth on
two sets of three lines. The lines and the spaces between them are 1.5~$\mu$m in the left
set, and 1.4~$\mu$m in the right set. A 3--$\mu$m--separation between the two sets 
allows for shadow--growth comparisons for systematically varied shadow geometries. 
The lower growth between the  lines in a particular set
is in contrast to the  higher growth between the two sets.
Sets of lines with decreasing linewidths and spacewidths are seen in Figs. 7b and c: going from left to right, the 
widths 
 decrease from  0.8  to  0.6~$\mu$m in Fig. 7b, and from 0.6 to 0.4~$\mu$m in Fig. 7. 
 There is no observable growth in the spaces of smaller widths,  and 
 very little growth on the sidewalls.

\begin{figure}[!ht]
\centering \psfull \epsfig{file=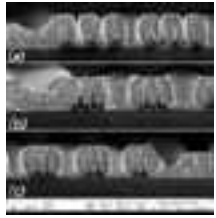}
 \caption{Cross--sectional SEM micrographs providing edge--on views (i.e.,
 at $90^\circ$ to the substrate normal) of the same
 film as in Fig. 6 for decreasing linewidths and spacewidths.
}
\end{figure}

Larger  
 separations of 3.5~$\mu$m (Fig. 7c) and 4.5~$\mu$m (Fig. 7b) can be seen
between adjacent  sets of lines. In these regions, growth on the bottom and on the sidewalls increases in direct proportion to the sidewall--to--sidewall separation. Nanowire arrays on the middle
line of each  set undergo an initial stage of lateral expansion, originating at the
photoresist corners and sidewalls, followed by a steady--state vertical growth. The  nanowire arrays 
on the middle lines maintain a nearly constant separation from the  
respective outer lines
for most of the 3.2--$\mu$m thickness of the film, with only slight rounding at the top edges. Also
notice how this separation distance decreases with the spacing between the photoresist lines.

Figures 8a--c are cross--sectional SEM micrographs of increasing magnification, 
showing the same two--section film described in Figs. 3, 6, and 7, 
 except that this portion of the film was grown on a 
 checkerboard pattern (2--D array) of photoresist.  The checkerboard
 dimensions are 1.5~$\mu$m~$\times$~1.5~$\mu$m steps and 
 wells. The chiral nanowires are 
 selectively deposited on top of the topographic features with no deposition occurring between the features,
 which is evident from Fig. 8c.  Thus,  2--D arrays
 of nanowire assemblies can be deposited over large areas (Fig. 8a), 
 with good uniformity (Fig. 8b), and with nearly vertical sidewalls (Fig. 8c). 

 \begin{figure}[!ht]
\centering \psfull \epsfig{file=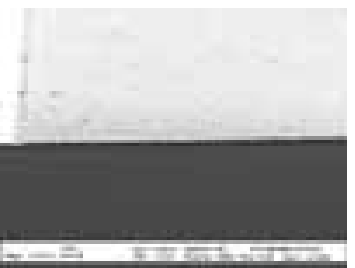}
 \epsfig{file=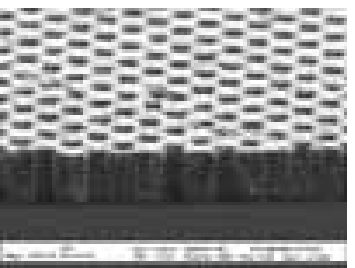} \\
(a) \hskip 1.25in (b)\\ \vskip0.2in
\epsfig{file=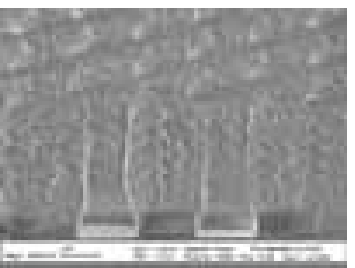}\\
(c)
\caption{Cross--sectional SEM micrographs of a two--section chiral STF
deposited on a checkerboard pattern (2--D array) of photoresist. 
}
\end{figure}

\section{Discussion}
The combination of large thickness ($>3$~$\mu$m), large--area uniformity (75~mm diameter), 
high growth rate (up to 0.4 $\mu$m/min) STFs with complex nanowire shapes (Figs. 2--4) 
on lithographically defined patterns (Figs. 5--8) has been achieved for the first time. 
Atomic self--shadowing due to oblique--angle deposition (at $75^\circ$
from the substrate normal, $\chiV=15^\circ$)  enables the nanowires to grow 
continuously, to change direction abruptly, and to maintain constant
cross--section, as seen in Fig. 2. 
By controlling the longer--range--shadowing of the collimated, directed vapor flux
by the photoresist pattern geometries, it has become possible to grow STFs 
selectively
on the top surfaces of both 1--D lines (Figs. 6 and 7) and 2--D 
checkerboard arrays (Fig. 8). The selectivity, as indicated by growth on the 
sidewalls of the photoresist features and the bare bottom surfaces, 
 is directly related to the feature spacing which, in turn, is related to the geometrical 
 shadowing distances.

The various morphological features of the STFs
 emanating from the photoresist sidewalls and the bottom (silicon)
 surfaces, as noted and described in the last section, can be understood in terms of simple 
 geometrical shadowing. The shadowing distance $\wS$ of a surface feature of sidewall height $\hS$
  due to an oblique angle flux, as shown in Fig. 9, is given by
 \begin{equation}
\wS = \gS \hS\, ,
 \end{equation}
where $\gS = 1/ \tan \chiV$ is a geometrical factor 
 and $\chiV$ is the angle between the average direction of the vapor flux and the
 substrate plane.
 This equation
 has been applied to atomic--level self--shadowing \c{4} and applies equally well to the sculptured
 nanowire assemblies presented here. 
 As
 $\chiV = 15\deg$ and $\hS = 0.8$~$\mu$m for all films
 in Figs. 2--8, $\gS = 3.73$. Therefore, the shadowing distance 
 when the vapor direction is perpendicular to the feature (e.g., photoresist line) is $\wS = 2.99$~$\mu$m. 
 This only describes shadowing of the bottom surface perpendicularly adjacent to the sidewall 
 of a $0.8$~$\mu$m--high feature, as in the case of 2--D nematic growth. If the separation
 of the features is less than $\wS$, the growth on the 
 sidewall of an adjacent feature (see Fig. 9) will coat only a fraction $\dSW$ of the sidewall
 from the top down; accordingly,
\begin{equation}
 \dSW = \frac{\wF}{\wS}\,,
 \end{equation}
where $\wF$ is the distance between the feature ledges.  
 For small feature separations, as studied here (see Figs. 7 and 8), 
 $\dSW$ is much smaller than unity. 
 
  \begin{figure}[!ht]
\centering \psfull \epsfig{file=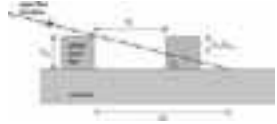}
\caption{Schematic of a two--dimensional well between two adjacent photoresist lines, and the
geometry of shadowing.
}
\end{figure}

With this simplistic model of shadowing behavior, it is possible to understand the 
 morphologies obtained in this study. In Fig. 5, for a single step
  where shadowing occurs only from 
 one direction, the width of zone 2 (deposition on the bottom surface) is about 3~$\mu$m which is close to the 
 shadowing distance. The morphology at the bottom on the left shows 
 columns slanted upward to the right in the direction where the vapor is arriving 
 unshadowed. On the right side of zone 2,  the columns becomes 
 more vertical, likely the result of shadowing from the faster growing film on the far 
 right that is growing un-shadowed in all directions. The transition zone 3 appears to 
 be related to the dynamic shadowing effects coupled with the change of deposition 
 from multidirectional (chiral) to bidirectional (chevronic) at which point the shadowing 
 geometry is that shown in Fig. 9. Similar arguments can be made for understanding the 
 morphology in Fig. 6a, though the transition from the shadowed to unshadowed zones is more 
 gradual, likely due to the multidirectional growth during the complete film deposition.

However, for many envisioned applications deposition will 
 be desired only on the top surface of a lithographically defined pattern.
This means that no vapor arrives at the bottom surface, as in the case of nematic growth 
 in the top section of the film in Fig.  5, as well as for chiral growth on submicron--wide 1--D
 line arrays (Figs. 7b and 7c) and  1.5~$\mu$m~$\times$~1.5~$\mu$m 2--D checkerboard arrays (Fig. 8).
 In general, the selectivity of growth on the top surface versus growth on the sidewalls or bottom
 surfaces for different pattern geometries and deposition sequences
 can be predicted. For instance, if $\wF$ is increased or $\hS$ is decreased, a decrease
 in $\chiV$ could be calculated for compensation.
 
 The shadowing geometry gets more complicated when the vapor flux is not
temporally unidirectional (inherent in the piecewise bidirectionality needed
for growing chevronic nanowires), but multidirectional (as for growing chiral nanowires).
  In the latter case as the substrate rotates  relative to the average direction
  of the incident vapor flux, growth will occur down the spaces between adjacent
  photoresist lines 
 casting an incident vapor flux that varies in angle and in position on the side--wall 
 and bottom surface.
 For substrate rotation at a constant angular velocity, as we undertook, the
 morphology of the deposited film will 
 vary.
 More detailed mathematical descriptions and/or simulations 
 of this complex growth will be needed if deposition in the channels between 
 features is important. Furthermore, such 
 a quantitative
 description will have to incorporate the dynamic effects of the advancing film growth--front in all directions. 
 
 The experimental evidence in Fig. 7 shows that the growth at the
 bottom surfaces between various 1--D arrays of photoresist
 lines (spaced 4.5, 3.5, 1.5, 1.4, 0.8, 0.6, 0.5, and 0.4~$\mu$m apart)
 decreased from about 25$\%$ of the unshadowed thickness of the
 4.5--$\mu$m--spacing (Fig. 7b), to
 about 12$\%$ for the  1.5--$\mu$m--spacing (Fig. 7a), and to essentially zero thickness
 for sidewall--to--sidewall separations of 0.8~$\mu$m
 and less (Figs. 7b and 7c). For the 2--D checkerboard array with
  1.5~$\mu$m~$\times$~1.5~$\mu$m~$\times$~0.8~$\mu$m shadowed wells (Fig. 8),
  there is no visible growth on the bottom surface, since the largest value
  of $\wS$ is $1.5\sqrt{2}$~$\mu$m along the checkerboard diagonal direction,
  which leads to $\dSW=0.71$. Thus, the growth will be just 
  $71\%$ of the sidewall growth in the worst case, with no
  growth
  predicted on the bottom surface. The sidewall growth leads to initial expansion
  of the nanowire assemblies (see Fig. 8c), followed by parallel, vertical growth,
  once the separation is about 1~$\mu$m. As the separation changes, so does the
  shadowing geometry. Therefore, the selectivity of growth on the pattern's top
  surface will improve as $\wF$ is reduced and/or $\chiV$ is decreased.

  \section{Concluding Remarks}
We have reported the fabrication of films comprising 
assemblies of sculptured SiO$_x$ nano\-wires grown on lithographically 
patterned substrates, thereby incorporating transverse architectures uniformly over large--area
substrates in STF technology.  The  density and morphology of the nanowire 
assemblies is dependent on the direction of the vapor flux and the 
local self--shadowing environment,
according to straightforward geometrical considerations.
We have shown that chiral nanowires can be selectively grown on patterned substrates. 
 The spacing of the array where selective growth is achievable is determined 
by the direction of the
vapor flux angle  and the height of the 
initial (i.e., substrate) topography.  The ability to sculpture the nanowires out of any material 
which can easily be deposited by a physical vapor deposition technique~---~coupled with 
micro-and macroscale 1-- and 2--dimensional topographic substrates~---~opens a 
whole new realm of photonic, fluidic and sensor devices \c{8,9,10,11,17}. Such applications
include nanowire assemblies for bio\-nano\-technology \c{18, 19}, as well as for
photonic bandgap engineering \c{20}.

{\small  \noindent {\bf Acknowledgement} This work was supported in part
by the Penn State Materials Research Science and Engineering Center.}

\end{document}